\RequirePackage{fix-cm}
\documentclass[twocolumn]{svjour3}          
\smartqed  
\usepackage{graphicx}
\usepackage{graphicx}
\usepackage{times}
\usepackage{latexsym}
\usepackage{todonotes}
\usepackage{listings}
\usepackage{subfig}
\usepackage{multirow}
\usepackage{tabularx}

\usepackage{rotating}
\usepackage{comment}
\usepackage{url}
\usepackage{subfig}
\usepackage{amssymb}
\usepackage{makecell}
\usepackage{enumitem}

\usepackage{scalerel}
\usepackage{tikz}
\usetikzlibrary{svg.path}

\definecolor{orcidlogocol}{HTML}{A6CE39}
\tikzset{
  orcidlogo/.pic={
    \fill[orcidlogocol] svg{M256,128c0,70.7-57.3,128-128,128C57.3,256,0,198.7,0,128C0,57.3,57.3,0,128,0C198.7,0,256,57.3,256,128z};
    \fill[white] svg{M86.3,186.2H70.9V79.1h15.4v48.4V186.2z}
                 svg{M108.9,79.1h41.6c39.6,0,57,28.3,57,53.6c0,27.5-21.5,53.6-56.8,53.6h-41.8V79.1z M124.3,172.4h24.5c34.9,0,42.9-26.5,42.9-39.7c0-21.5-13.7-39.7-43.7-39.7h-23.7V172.4z}
                 svg{M88.7,56.8c0,5.5-4.5,10.1-10.1,10.1c-5.6,0-10.1-4.6-10.1-10.1c0-5.6,4.5-10.1,10.1-10.1C84.2,46.7,88.7,51.3,88.7,56.8z};
  }
}

\newcommand\orcidicon[1]{\href{https://orcid.org/#1}{\mbox{\scalerel*{
\begin{tikzpicture}[yscale=-1,transform shape]
\pic{orcidlogo};
\end{tikzpicture}
}{|}}}}

\usepackage{hyperref}
\usepackage[layout=footnote]{fixme}
\graphicspath{ {./img/} }

\newcolumntype{Y}{>{\centering\arraybackslash}X}

\begin{document}

\title{Analysing the Requirements for an Open Research Knowledge Graph: Use Cases, Quality Requirements and Construction Strategies}
\titlerunning{Requirements Analysis for an Open Research Knowledge Graph}        

\author{
Arthur Brack\textsuperscript{1,2}\orcidicon{0000-0002-1428-5348}
\and Anett Hoppe\textsuperscript{1}\orcidicon{0000-0002-1452-9509}
\and Markus Stocker\textsuperscript{1}\orcidicon{0000-0001-5492-3212}
\and S\"oren Auer\textsuperscript{1,2}\orcidicon{0000-0002-0698-2864}
\and Ralph Ewerth\textsuperscript{1,2}\orcidicon{0000-0003-0918-6297}}


\authorrunning{Brack et al.} 

\institute{Arthur Brack  \\
              \email{arthur.brack@tib.eu}            \\
           Anett Hoppe  \\
              \email{anett.hoppe@tib.eu}  \\
           Markus Stocker  \\
              \email{markus.stocker@tib.eu}  \\
           S\"oren Auer  \\
              \email{auer@tib.eu}    \\            
           Ralph Ewerth  \\
              \email{ralph.ewerth@tib.eu}                              
            \at
               \textsuperscript{1}TIB -- Leibniz Information Centre for Science and Technology, Hannover, Germany         
            \at
               \textsuperscript{2}L3S Research Center, Leibniz University, Hannover, Germany
}

\date{Received: date / Accepted: date}

\maketitle

\begin{abstract}
Current science communication has a number of drawbacks and bottlenecks which have been subject of discussion lately: Among others, the rising number of published articles makes it nearly impossible to get a full over\-view of the state of the art in a certain field, or reproducibility is hampered by fixed-length, document-based publications which normally cannot cover all details of a research work. Recently, several initiatives have proposed knowledge graphs (KG) for organising scientific information as a solution to many of the current issues. The focus of these proposals is, however, usually restricted to very specific use cases. In this paper, we aim to transcend this limited perspective and present a comprehensive analysis of requirements for an Open Research Knowledge Graph (ORKG) by (a) collecting and reviewing daily core tasks of a scientist, (b) establishing their consequential requirements for a KG-based system, (c) identifying overlaps and specificities, and their coverage in current solutions. As a result, we map necessary and desirable requirements for successful KG-based science communication, derive implications, and outline possible solutions.


\end{abstract}

\section{Introduction}
Today's scholarly communication is a document-centred process and as such, rather inefficient. Scientists spend considerable time in finding, reading and reproducing research results from PDF files consisting of static text, tables, and figures. The explosion in the number of published articles~\cite{bornmann15growth} aggravates this situation further: It gets harder and harder to stay on top of current research, that is to find relevant works, compare and reproduce them and, later on, to make one's own contribution known for its quality. 

Some of the available infrastructures in the research eco\-system already use \emph{knowledge graphs} (KG)\footnote{Acknowledging that knowledge graph is vaguely defined, we adopt the following definition: A \emph{knowledge graph} (KG) consists of (1) an \emph{ontology} describing a conceptual model (e.g. with classes and relation types), and (2) the corresponding \emph{instance data} (e.g. objects, literals, and \textless subject, predicate, object\textgreater-triplets) following the constraints posed by the ontology (e.g. instance-of relations).
The construction of a KG involves \emph{ontology design} and \emph{population} with instances.} 
to enhance their services. 
Academic search engines, for instance, such as \textit{Microsoft Academic Knowledge Graph}~\cite{Farber2019TheMA} or \textit{Literature Graph} \cite{Ammar2018ConstructionOT} utilise metadata-based graph structures which link research articles based on citations, shared authors, venues and keywords. 

Recently, initiatives have promoted the usage of KGs in science communication, but on a deeper, semantic level \cite{Auer19,Hars2003StructureOS,Jaradeh2019OpenRK,Manghi_Bardi_Atzori_Baglioni_Manola_Schirrwagen_Principe_2019,Oelen20,pertsas2017scholarly,Zhang2019}. 
They envision the transformation of the dominant document-centred knowledge exchange to know\-ledge-based information flows by representing and expressing knowledge through semantically rich, interlinked KGs.
Indeed, they argue that a shared structured representation of scientific knowledge has the potential to alleviate some of the science communication's current issues: Relevant research could be easier to find, comparison tables automatically compiled, own insights rapidly placed in the current ecosystem. 
Such a powerful data structure could, more than the current document-based system, also encourage the interconnection of research artefacts such as datasets and source code much more than current approaches (like Digital Object Identifier (DOI) references etc.); allowing for easier reproducibility and comparison. 
To come closer to the vision of know\-ledge-based information flows, research articles should be enriched and interconnected through machine-interpretable semantic content. 
The usage of Papers With Code \cite{PWC} in the machine learning community and Jaradeh et al.'s study~\cite{Jaradeh2019OpenRK} indicate that authors are also willing to contribute structured descriptions of their research articles.

The work of a researcher is manifold, but current proposals usually focus on a specific use case (e.g. the aforementioned examples focus on enhancing academic search). In this paper, we present a detailed analysis of common literature-related tasks in a scientist's daily life and analyse (a) how they could be supported by an ORKG, (b) what requirements result for the design of (b1) the KG and (b2) the surrounding system, (c) how different use cases overlap in their requirements and can benefit from each other. Our analysis is led by the following research questions: 

\begin{enumerate}
    \item Which use cases should be supported by an ORKG?
    \begin{enumerate}
        \item Which user interfaces are necessary?
        \item Which machine interfaces are necessary? 
    \end{enumerate}
    \item What requirements can be defined for the underlying ontologies to support these use cases? 
    \begin{enumerate}
        \item Which granularity of information is needed?
        \item To what degree is domain specialisation needed?
    \end{enumerate}
    \item What requirements can be defined for the instance data in context of the respective use cases?
    \begin{enumerate}
        \item Which completeness is sufficient for the instance data?
        \item Which correctness is sufficient for the instance data? 
        \item Which approaches (human vs. machine) are suitable to populate the ORKG? 
    \end{enumerate}    
\end{enumerate}
We follow the design science research (DSR) methodology \cite{Hevner2004DesignSI}. In this study, we focus on the first phase of DSR and conduct a requirements analysis. The objective is to chart necessary (and desirable) requirements for successful KG-based science communication, and, consequently, provide a map for future research.  

Compared to our paper at the 24th International Conference on Theory and Practice of Digital Libraries 2020  \cite{Brack2020RequirementsAnalysisORKG}, this journal paper has been modified and extended as follows: 
The related work section is updated and extended with the new sections \emph{Quality of knowledge graphs} and \emph{Systematic literature reviews}. The new Appendix~\ref{sec:comparisons_information_extraction} contains 
comparative overviews of datasets for research knowledge graph population tasks such as sentence classification, relation extraction, and concept extraction. To be consistent with terminology in related work, we use the term ``completeness'' instead of ``coverage'' and ``correctness'' instead of ``quality''. The requirements analysis in Section 3 is revised and contains more details with more justifications for the posed requirements and approaches.

The remainder of the paper is organised as follows. Section 2 summarises related work on research knowledge graphs, scientific ontologies, KG construction, data quality requirements, and systematic literature reviews. The requirements analysis is presented in Section 3, while Section 4 discusses implications and possible approaches for ORKG construction. Finally, Section 5 concludes the requirements analysis and outlines areas of future work. Appendix~\ref{sec:comparisons_information_extraction} contains comparative overviews for the tasks of sentence classification, relation extraction, and concept extraction.

\section{Related work}
\label{sec:related_work}
This section gives a brief overview of 
(a) existing research KGs,
(b) ontologies for scholarly knowledge,
(c) approaches for KG construction,
(d) quality dimensions of KGs, and
(e) processes in systematic literature reviews.

\subsection{Research knowledge graphs}

Academic search engines (e.g. Google Scholar, Microsoft Academic, SemanticScholar) exploit graph structures such as the Microsoft Academic Knowledge Graph~\cite{Farber2019TheMA}, SciGraph \cite{Yaman2019InterlinkingSA}, the Literature Graph~\cite{Ammar2018ConstructionOT}, 
or the Semantic Scholar Open Research Corpus (S2ORC) \cite{Lo2020}.
These graphs interlink research articles through metadata, e.g. citations, authors, affiliations, grants, journals, or keywords.

To help reproduce research results, initiatives such as Research Graph~\cite{Amir2017ResearchGB}, Research Objects~\cite{Bechhofer2010WhyLD} and OpenAIRE \cite{Manghi_Bardi_Atzori_Baglioni_Manola_Schirrwagen_Principe_2019}
interlink research articles with research artefacts such as data\-sets, source code, software, and video presentations.
Scholarly Link Exchange (Scholix)~\cite{Burton_2017} aims to create a standardised ecosystem to collect and exchange links between research artefacts and literature.

Some approaches connect articles at a more semantic level:
Papers With Code \cite{PWC} is a community-driven effort to supplement machine learning articles with tasks, source code and evaluation results to construct leaderboards.
Ammar et al.~\cite{Ammar2018ConstructionOT} link entity mentions in abstracts with DBpedia~\cite{Lehmann2015DBpediaA} and Unified Medical Language System (UMLS) \cite{Bodenreider2004TheUM}, and Cohan et al.~\cite{Cohan2019StructuralSF} extend the citation graph with citation intents (e.g. citation as background or used method).

Various scholarly applications benefit from semantic content representation, e.g. academic search engines by exploiting general-purpose KGs~\cite{Xiong2017ExplicitSR},
and graph-based research paper recommendation systems~\cite{Beel2015ResearchpaperRS} by utilising citation graphs and mentioned entities.
However, the coverage of science-specific concepts in general-purpose KGs is rather low \cite{Ammar2018ConstructionOT},
e.g. the task ``geolocation estimation of photos'' from Computer Vision is neither present in Wikipedia nor in the Computer Science Ontology (CSO) \cite{Salatino2019TheCS}.

\subsection{Scientific ontologies}

Various ontologies have been proposed to model metadata such as bibliographic resources and citations~\cite{Peroni2012FaBiOAC}. 
Iniesta and Corcho~\cite{IniestaSurveyOntologies} reviewed ontologies to describe scholarly articles.
In the following, we describe some ontologies that conceptualise the semantic content in research articles. 

Several ontologies focus on rhetorical~\cite{Waard2006TheAF,Groza2007SALTS,Constantin2016TheDC} (e.g. Background, Methods, Results, Conclusion), 
argumentative \cite{teufel2009towards,Liakata2010CorporaFT} (e.g. claims, contrastive and comparative statements about other work) or 
activity-based structure~\cite{pertsas2017scholarly} (e.g. sequence of research activities) of research articles.
Others describe scholarly knowledge with linked entities such as problem, method, theory, statement~\cite{Hars2003StructureOS,Brodaric2008SKIingWD}, or focus on the main research findings and characteristics of research articles described in surveys with concepts such as problems, approaches, implementations, and evaluations~\cite{Fathalla2017TowardsAK,Vahdati2019SemanticRO}.

Various domain-specific ontologies exist, for instance, 
mathematics~\cite{Lange2013OntologiesAL} (e.g. definitions, assertions, proofs), 
machine learning~\cite{Klampanos2018ANNETTOAO,Mesbah2017SemanticAO} (e.g. dataset, metric, model, experiment), and
physics ~\cite{Say2020} (e.g. formation, model, observation).
The EXPeriments Ontology (EXPO) is a core ontology for scientific experiments that conceptualises experimental design, methodology, and results~\cite{Soldatova2006AnOO}.

Taxonomies for domain-specific research areas support the characterisation and exploration of a research field.
Sala\-tino et al.~\cite{Salatino2019TheCS} give an overview, e.g. Medical Subject Heading (MeSH), Physics Subject
Headings (PhySH), Computer Science Ontology (CSO).
Gene Ontology~\cite{Carbon2019} and Chemical Entities of Biological Interest (CheBi)~\cite{Degtyarenko2007ChEBIAD} are KGs for genes and molecular entities.

\subsection{Construction of knowledge graphs}
\label{sec:construction_KG}

Nickel et al.~\cite{Nickel16} classify KG construction methods into four groups: (1) curated approaches, i.e. triples created  manually by a closed group of experts, (2) collaborative approaches, i.e. triples created manually by an open group of volunteers, (3) automated semi-structured approaches, i.e. triples extracted automatically from semi-structured text via hand-crafted rules, and (4) automated unstructured approaches, i.e. triples are extracted automatically from unstructured text.

\subsubsection{Manual approaches}

WikiData~\cite{vrandevcic2014wikidata} is one of the most popular KGs with semantically structured, encyclopaedic knowledge curated manually by a community. 
As of January 2021, WikiData comprises 92M entities curated by almost 27.000 active contributors.
The community also maintains a taxonomy of categories and "infoboxes" which define common properties of certain entity types.
Furthermore, Papers With Code \cite{PWC} is a community-driven effort to interlink machine learning articles with tasks, source code and evaluation results.
KGs such as Gene Ontology~\cite{Carbon2019} or Wordnet~\cite{Fellbaum2000WordNetA} are curated by domain experts.
Research article submission portals such as EasyChair~(\url{https://wwww.easychair.org/}) enforce the authors to provide machine-readable metadata.
Librarians and publishers tag new articles with keywords and subjects~\cite{Yaman2019InterlinkingSA}.
Virtual research environments enable the execution of data analysis on interoperable infrastructure and store the data and results in KGs~\cite{Stocker2018TowardsRI}.

\subsubsection{Automated Approaches}

\paragraph{Automatic KG construction from text:}
Petasis et al.~\cite{PetasisKPKZ11} pre\-sent 
a review on \emph{ontology learning}, that is ontology creation from text, while Lubani et al.\cite{LubaniNM19} review \emph{ontology population systems}. 
Pajura and Singh~\cite{Pujara2018MiningKG} give an overview of the involved tasks for \emph{KG population}:
(a) \emph{information extraction} to extract a graph from text with \emph{entity extraction} and \emph{relation extraction}, and (b) \emph{graph construction} to clean and complete the extracted graph, as it is usually ambiguous, incomplete and inconsistent.  
\emph{Coreference resolution}~\cite{Brack2021Coref,Luan2018MultiTaskIO} clusters different mentions of the same entity in text and \emph{entity linking}~\cite{Kolitsas2018EndtoEndNE} maps mentions in text to entities in the KG.
\emph{Entity resolution}~\cite{TALBURT201139} identifies objects in the KG that refer to the same underlying entity.
For \emph{taxonomy population}, Salatino et al.~\cite{Salatino2019TheCS} provide an overview of methods based on rule-based natural language processing (NLP), clustering and statistical methods.

The Computer Science Ontology (CSO) has been automatically populated from research articles~\cite{Salatino2019TheCS}.
The AI-KG was automatically generated 
from 333,000 research papers in the artificial intelligence (AI) domain~\cite{Dessi2020AIKG}. It contains five entity types (tasks,
methods, metrics, materials, others) linked by 27 relations types.
Kannan et al.~\cite{Kannan20} create  
a multimodal KG for deep learning papers from text and images and the corresponding source code.
Brack et al.~\cite{Brack2021Coref} generate 
a KG for 10 different science domains with the concept types material, method, process, and data.
Zhang et al.~\cite{Zhang2019} suggest 
a rule-based approach to mine research problems and proposed solutions from research papers.

\paragraph{Information extraction from scientific text:}
\label{sec:scientific_information_extraction}
Information extraction is the first step in the automatic KG population pipe\-line.
Nasar~et~al.~\cite{Nasar2018InformationEF} 
survey methods on information extraction from scientific text.
Beltagy~et~al.~\cite{Beltagy2019SciBERTPC} present benchmarks for several scientific datasets and Peng et al.~\cite{PengYL19} especially for the biomedical domain.
Appendix~\ref{sec:comparisons_information_extraction} presents 
comparative overviews of datasets for the tasks sentence classification, relation extraction, and concept extraction, respectively, in research papers.

There are datasets which are annotated at \emph{sentence level} for several domains, e.g. biomedical~\cite{Dernoncourt2017PubMed2R,Kim2011AutomaticCO}, computer graphics~\cite{Fisas2015OnTD}, computer science~\cite{Cohan2019PretrainedLM}, chemistry and computational linguistics~\cite{teufel2009towards}, or algorithmic metadata~\cite{Safder2020}. 
They cover either only 
abstracts \cite{Dernoncourt2017PubMed2R,Kim2011AutomaticCO,Cohan2019PretrainedLM} 
or full articles \cite{Fisas2015OnTD,Liakata2010CorporaFT,Safder2020,teufel2009towards}.
The datasets differentiate between five and twelve concept classes (e.g. Background, Objective, Results).
Machine learning approaches for data\-sets consisting of abstracts achieve an F1 score ranging from 66\% to 92\% and for datasets with full papers F1 scores range from 51\% to 78\% (see Table~\ref{tab:comparison_sentence_classification}).

More recent corpora, annotated at \emph{phrasal level}, aim at constructing a fine-grained KG from scholarly abstracts with the tasks of concept extraction~\cite{augenstein2017semeval,friedrich-etal-2020-sofc,Luan2018MultiTaskIO,Brack2020DomainindependentEO,handschuh2014acl}, 
binary relation extraction~\cite{Luan2018MultiTaskIO,gabor2018semeval,augenstein2017semeval},
n-ary relation extraction~\cite{Kardas2020-axcell,Jain2020,Jia2019},
and coreference resolution~\cite{Brack2021Coref,CohenLCBBPVPH17Craft,Luan2018MultiTaskIO}.
They cover several domains, 
e.g. 
material sciences~\cite{friedrich-etal-2020-sofc};
computational linguistics~\cite{gabor2018semeval,handschuh2014acl};
computer science, material sciences, and physics~\cite{augenstein2017semeval};
machine learning~\cite{Luan2018MultiTaskIO}; 
biomedicine~\cite{CohenLCBBPVPH17Craft,Jia2019,Kringelum2016}; 
or a set of ten scientific, technical and medical domains~\cite{Brack2020DomainindependentEO,Brack2021Coref,DSouza2020STEM}.
The datasets differentiate between four to seven concept classes (like Task, Method, Tool) 
and between two to seven binary relation types (like used-for, part-of, evaluate-for).
The extraction of n-ary relations involves extraction of relations among multiple concepts such as drug-gene-mutation interactions in medicine \cite{Jia2019}, experiments related to solid oxide fuel cells with involved material and measurement conditions in material sciences \cite{friedrich-etal-2020-sofc}, or task-dataset-metric-score tuples for leaderboard construction for machine learning tasks \cite{Kardas2020-axcell}. 

Approaches for concept extraction achieve F1 scores ranging from 56.6\% to 96.9\% (see Table~\ref{tab:comparison_concept_extraction}), for coreference resolution F1 scores range from 46.0\% to 61.4\%~\cite{Brack2021Coref,CohenLCBBPVPH17Craft,Luan2018MultiTaskIO}, and for binary relation extraction from 28.0\% to 83.6\% (see Table~\ref{tab:comparison_relation_extraction}).
The task of n-ary relation extraction with an F1 score from 28.7\% to 56.4\%~\cite{Jia2019,Kardas2020-axcell} is especially challenging, since such relationships usually span beyond sentences or even sections and thus, machine learning models require an understanding of the whole document. 
The inter-coder agreement for the task of concept extraction ranges from 0.6 to 0.96 (Table~\ref{tab:comparison_concept_extraction}), for relation extraction from 0.6 to 0.9 (see also Table~\ref{tab:comparison_relation_extraction}), while for coreference resolution the value of 0.68 was reported in two different studies~\cite{Brack2021Coref,Luan2018MultiTaskIO}.
The results suggest that these tasks are not only difficult for machines but also for humans in most cases.

\subsection{Quality of knowledge graphs}
\label{sec:data_quality}
KGs may contain billions of machine-readable facts about the world or a certain domain. However, do the KGs have also an appropriate quality?
Data quality (DQ) is defined as \emph{fitness for use by a data consumer}~\cite{Wang96}. Thus, to evaluate data quality, it is important to know the needs of the data consumer since, in the end, the consumer judges whether or not a product is fit for use.
Wang et al.~\cite{Wang96} propose a data quality evaluation framework for information systems consisting of 15 dimensions grouped into four categories, i.e.:
\begin{enumerate}
\item \emph{Intrinsic DQ}: accuracy, objectivity, believability, and reputation.
\item \emph{Contextual DQ}: value-added, relevancy, timeliness, completeness, and an appropriate amount of data.
\item \emph{Representational DQ}: interpretability, ease of understanding, representational consistency, and concise representation.
\item \emph{Accessibility DQ}: accessibility and access security.
\end{enumerate}

Bizer~\cite{Bizer07} and Zaveri~\cite{Zaveri16} propose further dimensions for the Linked Data context like consistency, verifiability, offensiveness, licensing and interlinking. 
Pipino et al.~\cite{Pipino02} subdivide completeness into \emph{schema completeness}, i.e. the extent to which classes and relations are missing in the ontology to support a certain use, \emph{column completeness} (also known as \emph{Partial Closed World Assumption}~\cite{Galarraga13}), i.e. the extent to which facts are not missing, and \emph{population completeness}, i.e. the extent to which instances for a certain class are missing. F\"arber et al.~\cite{Farber18} comprehensively evaluate and compare the data quality of popular KGs (e.g. DBpedia, Freebase, Wikidata, YAGO) using such dimensions.

To evaluate the correctness of instance data (also known as \emph{precision}), the facts in the KG have to be compared against a ground truth. For that, humans annotate a set of facts as true or false.
YAGO found to be 95\% correct~\cite{Yago07}.
The automatically populated AI-KG has a precision of 79\%~\cite{Dessi2020AIKG} .
The KG automatically populated by the Never-Ending Language Learner (NELL) has a precision of 74\%~\cite{Carlson10}.

To evaluate the \emph{completeness of instance data} (also known as \emph{coverage and recall}), small collections of ground-truth capturing \emph{all} knowledge for a certain ontology is necessary, that are usually difficult to obtain~\cite{Weikum2020MachineKnowledge}.
However, some studies estimate the completeness of several KGs.
Galarrage et al.~\cite{Galarraga17} suggest a rule mining approach to predict missing facts.
In Freebase~\cite{BollackerEPST08} 71\% of people have an unknown place of birth, and 75\% have an unknown nationality~\cite{Dong2014}.
Suchanek et al.~\cite{SuchanekGA11} report that 69\%-99\% of instances in popular KGs (e.g. YAGO, DBPedia) do not have at least one property that other instances of the same class have.
The AI-KG has a recall of 81.2\%~\cite{Dessi2020AIKG}.

\subsection{Systematic literature reviews}
\label{sec:literature_reviews}
\begin{figure*}[t!]
    \center{\includegraphics[width=0.8\linewidth]
        {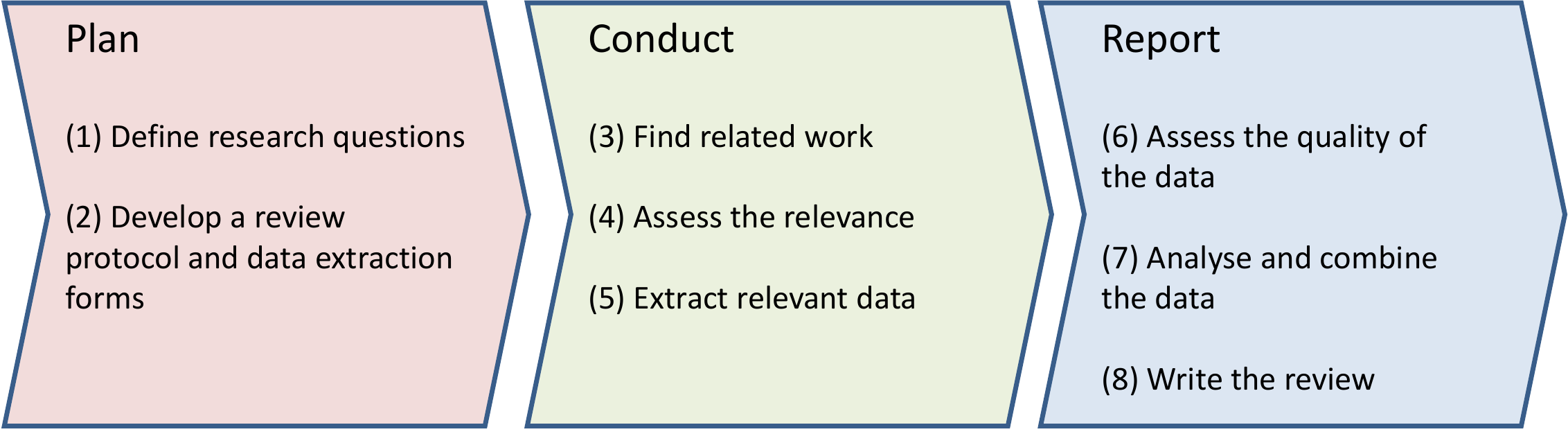}}
    \caption{Activities within a systematic literature review}
    \label{fig:steps_slr}
\end{figure*}

Literature reviews are one of the main tasks of researchers, since a
clear identification of a contribution to the present scholarly knowledge is a crucial step in scientific work~\cite{Hevner2004DesignSI}. This requires a comprehensive elaboration of the present scholarly knowledge for a certain research question. 
Furthermore, systematic literature reviews help to identify research gaps and to position new research activities~\cite{Kitchenham07guidelinesfor}.

A literature review can be conducted systematically or in a non-systematic, narrative way.
Following Fink's~\cite{fink2014conducting} definition, a systematic literature review is \textit{``a systematic, explicit, comprehensive, and reproducible method identifying, evaluating, and synthesising the existing body of completed and recorded work''}. 
Guidelines for systematic literature reviews have been suggested for several scientific disciplines, e.g. for software engineering \cite{Kitchenham07guidelinesfor}, for information systems \cite{Okoli2015AGT} and for health sciences \cite{fink2014conducting}. 
A systematic literature review consists typically of the activities depicted in Figure~\ref{fig:steps_slr} subdivided into the phases \emph{plan}, \emph{conduct}, and \emph{report}. The activities may differ in detail for the specific scientific domains~\cite{Kitchenham07guidelinesfor,Okoli2015AGT,fink2014conducting}. 
In particular, a \emph{data extraction form} defines which data has to be extracted from the reviewed papers. Data extraction requirements vary from review to review so that the form is tailored to the specific research questions investigated in the review.

\section{Requirements analysis}
As the discussion of related work reveals, 
existing knowledge graphs for research information focus on specific use cases (e.g. improve search engines, help to reproduce research results) and mainly manage metadata and research artefacts about articles.
We envision a KG in which research articles are linked through a deep semantic representation of their content to enable further use cases. In the following, we formulate the problem statement and describe our research method. This motivates our use case analysis in Section \ref{sec:use_cases}, from which we derive requirements for an ORKG. 

\paragraph{Problem statement:}
Scholarly knowledge is very heterogeneous and diverse. 
Therefore, an ontology that conceptualises scholarly knowledge comprehensively does not exist.
Besides, due to the complexity of the task, the population of comprehensive ontologies requires domain and ontology experts. 
Current automatic approaches can only populate rather simple ontologies and achieve moderate accuracy (see Section~\ref{sec:construction_KG} and Appendix~\ref{sec:comparisons_information_extraction}).
\emph{On the one hand, we desire an ontology that can comprehensively capture scholarly knowledge, and instance data with high correctness and completeness. On the other hand, we are faced with a ``knowledge acquisition bottleneck''.}

\paragraph{Research method:}
To illuminate the problem statement, we perform a \emph{requirements analysis}.
We follow the \emph{design science research (DSR)} methodology~\cite{Horvth2007ComparisonOT,Braun2015ProposalFR}. The requirements analysis is a central phase in DSR, as it is
the basis for design decisions and selection of methods to construct effective solutions systematically~\cite{Braun2015ProposalFR}. 
The objective of DSR in general is the innovative, rigorous and relevant design of information systems for solving important business problems, or the improvement of existing solutions~\cite{Braun2015ProposalFR,Hevner2004DesignSI}.

To elicit requirements, we studied guidelines for (a) systematic literature reviews (see Section~\ref{sec:literature_reviews}), (b) data quality requirements for information systems (see Section~\ref{sec:data_quality}), 
and (c) interviewed members of the ORKG and Visual Analytics team at TIB\footnote{\url{https://projects.tib.eu/orkg/project/team/}, \url{https://www.tib.eu/en/research-development/visual-analytics/staff}}, who are software engineers and researchers in the field of computer science and environmental sciences.
Based on the requirements, we elaborate possible approaches to construct an ORKG, which were identified through a literature review (see Section~\ref{sec:construction_KG}).
To verify our assumptions on the presented requirements and approaches, ORKG and Visual Analytics team members reviewed them in an iterative refinement process.

\subsection{Overview of the use cases}
\label{sec:use_cases}

\begin{figure*}[tb]
    \center{\includegraphics[width=\linewidth]
        {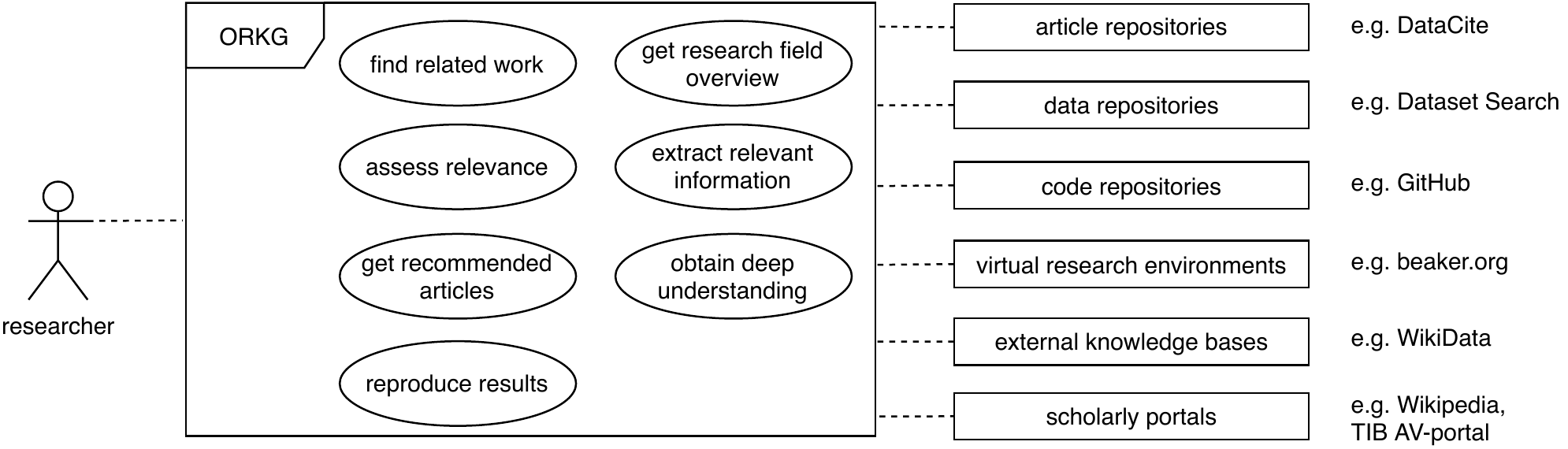}}
    \caption{UML use case diagram for the main use cases between the actor researcher, an Open Research Knowledge Graph (ORKG), and external systems.}
    \label{fig:use_cases}
\end{figure*}

We define functional requirements with use cases which are a popular technique in software engineering~\cite{Booch2015TheUM}. 
A use case describes the interaction between a user and the system from the \emph{user's perspective} to achieve a certain goal. 
Furthermore, a use case introduces 
a motivating scenario to guide the design of a supporting ontology and the use case analysis helps to figure out which kind of information is necessary~\cite{Degbelo2017ASO}. 

There are many use cases (e.g. literature reviews, plagiarism detection, peer reviewer suggestion) and several stakeholders (e.g. researchers, librarians, peer reviewers, practitioners) that may benefit from an ORKG.
Ngyuen et al.~\cite{Nguyen20} discuss some research-related tasks of scientists for information foraging at a broader level.
In this study, we focus on use cases that support \emph{researchers} (a) conducting literature reviews (see also Section~\ref{sec:literature_reviews}), (b) obtaining a deep understanding of a research article and (c) reproducing research results. A full discussion of all possible use cases of graph-based knowledge management systems in the research environment is far beyond the scope of this article. With the chosen focus, we hope to cover the most frequent, literature-oriented tasks of scientists.

Figure~\ref{fig:use_cases} depicts the main identified use cases, which are described briefly in the following. 
Please note that we focus on how \emph{semantic content} can improve these use cases and not further metadata.

\paragraph{Get research field overview:}
Survey articles provide an over\-view of a particular research field, e.g. a certain research problem or a family of approaches. 
The results in such surveys are sometimes summarised in structured and comparative tables (an approach usually followed in domains such as computer science, but not as systematically practised in other fields). 
However, once survey articles are published they are no longer updated. 
Moreover, they usually represent only the perspective of the authors, i.e. very few researchers of the field.
To support researchers to obtain an up-to-date overview of a research field, the system should maintain such surveys in a structured way, and allow for dynamics and evolution.
A researcher interested in such an overview should be able to search or to browse the desired research field in a user interface for ORKG access.
Then, the system should retrieve  
related articles and available overviews, e.g. in a table or a leaderboard chart. 

While an ORKG user interface should allow for showing tabular leaderboards or other visual representations, the backend should semantically represent information to allow for the exploitation of overlaps in conceptualisations between research problems or fields.
Furthermore, faceted drill-down methods based on the properties of semantic descriptions of research approaches could empower researchers to quickly filter and zoom into the most relevant literature.

\paragraph{Find related work:}
Finding relevant research articles is a daily core activity of researchers.
The primary goal of this use case is to find research articles which are relevant to a certain research question. A broad research question is often broken down into smaller, more specific sub-questions which are then converted to search queries~\cite{fink2014conducting}. For instance, in this paper, we explored the following sub-questions: 
(a) \emph{Which ontologies do exist to represent scholarly knowledge?}
(b) \emph{Which scientific knowledge graphs do exist and which information do they contain?}
(c) \emph{Which datasets do exist for scientific information extraction?}
(d) \emph{What are current state-of-the-art methods for scientific information extraction?}
(e) \emph{Which approaches do exist to construct a knowledge graph?}

An ORKG should support the answering of queries related to such questions, which can be fine-grained or broad search intents. 
Preferably, the system should support natural language queries as approached by semantic search and question answering engines~\cite{Balog2018EntityOrientedS}. The system has to return a set of relevant articles.

\paragraph{Assess relevance:}
Given a set of relevant articles the researcher has to assess whether the articles match the criteria of interest.
Usually researchers skim through the title and abstract. Often, also the introduction and conclusions have to be considered, which is cumbersome and time-consuming. If only the most important paragraphs in the article are presented to the researcher in a structured way, this process can be boosted. Such information snippets might include, for instance, text passages that describe the problem tackled in the research work, the main contributions, the employed methods or materials, or the yielded results. 

\begin{figure*}[htb]
    \center{\includegraphics[width=1.0\linewidth]
        {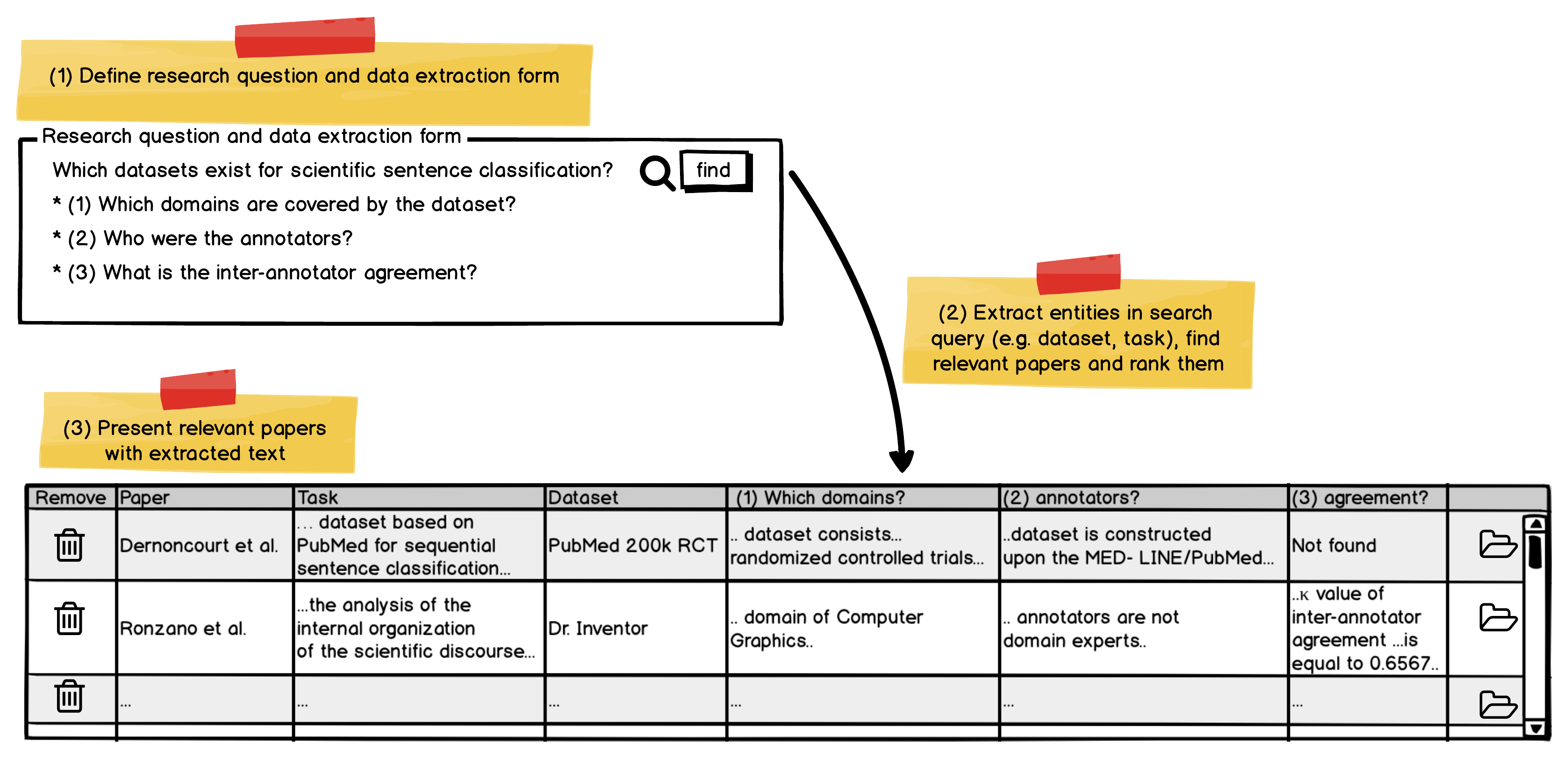}}
    \caption{An example research questions with a corresponding data extraction form, and the extracted text passages from relevant research articles for the respective (data extraction form) fields presented in a tabular form.}
    \label{fig:gui_mockup}
\end{figure*}

\paragraph{Extract relevant information:}
To tackle a particular research question, the researcher has to extract relevant information from research articles.
In a systematic literature review, the information to be extracted can be defined through a \emph{data extraction form} (see Section~\ref{sec:literature_reviews}).
Such extracted information is usually compiled in written text or comparison tables in a related work section or survey articles.
For instance, for the question \emph{"Which datasets do exist for scientific sentence classification?"} a researcher who focuses on a new annotation study could be interested in (a) domains covered by the dataset and (b) the inter-coder agreement (see Table~\ref{tab:comparison_sentence_classification} as an example). 
Another researcher might follow the same question but focusing on machine learning, and thus could be more interested in (c) evaluation results and (d) feature types used. 

The system should support the researcher with tailored information extraction from a set of research articles: (1) the researcher defines a data extraction form as proposed in systematic literature reviews (e.g. the fields (a)-(d)), and (2) the system presents the extracted information as suggestions for the corresponding data extraction form and articles in a comparative table.
Figure~\ref{fig:gui_mockup} illustrates a data extraction form with corresponding fields in form of questions, and a possible approach to visualise the extracted text passages from the articles for the respective fields in a tabular form.

\paragraph{Get recommended articles:}
When the researcher focuses on a particular article, further related articles could be recommended by the system utilising an ORKG, for instance, articles that address the same research problem or apply similar methods.

\paragraph{Obtain deep understanding:}
The system should help the researcher to obtain a deep understanding of a research article (e.g. equations, algorithms, diagrams, datasets). 
For this purpose, the system should connect the article with artefacts such as conference videos, presentations, source code, datasets, etc., and visualise the artefacts appropriately.
Also text passages can be linked, e.g. explanations of methods in Wikipedia, source code snippets of an algorithm implementation, or equations described in the article.

\paragraph{Reproduce results:}
The system should offer 
researchers links to all necessary artefacts to help to reproduce research results, e.g. datasets, source code, virtual research environments, materials describing the study, etc.
Furthermore, the system should maintain semantic descriptions of domain-specific and standardised evaluation protocols and guidelines such as in machine learning reproducibility checklists \cite{Pineau20} and bioassays in the medical domain.

\subsection{Knowledge graph requirements}
As outlined in Section~\ref{sec:data_quality}, data quality requirements should be considered within the context of a particular use case (``fitness for use'').
In this section, we first describe dimensions we used to define non-functional requirements for an ORKG. 
Then, we discuss these requirements within the context of our identified use cases.

\subsubsection{Dimensions for KG requirements}

In the following, we describe the dimensions that we use to define the requirements for ontology design and instance data. We selected these dimensions since we assume that they are most relevant and also challenging to construct an ORKG with appropriate data to support the various use cases.

For \emph{ontology design}, i.e. how comprehensively should an ontology conceptualise scholarly knowledge to support a certain use case, we use the following dimensions:
\begin{itemize}

\item[A)] \emph{Domain specialisation of the ontology:} 
How domain-specific should the concepts and relation types be in the ontology?
An ontology with \emph{high domain specialisation} targets a specific (sub-)domain and uses domain-specific terms. 
An ontology with \emph{low domain specialisation} targets a broad range of domains and uses rather domain-independent terms.
For instance, various ontologies (e.g. ~\cite{pertsas2017scholarly,Brack2020DomainindependentEO}) propose domain independent concepts (e.g. Process, Method, Material).
In contrast, Klampanos et al.~\cite{Klampanos2018ANNETTOAO} present a very domain-specific ontology for artificial neural networks.

\item[B)] \emph{Granularity of the ontology:} 
Which granularity of the ontology is required to conceptualise scholarly knowledge?
An ontology with \emph{high granularity} conceptualises scholarly knowledge with a lot of classes that have very detailed and a lot of fine-grained properties and relations.
An ontology with a \emph{low granularity} has only a few classes and relation types.
For instance, the annotation schemes for scientific corpora (see Section~\ref{sec:construction_KG}) have a rather low granularity, as they do not have more than 10 classes and 10 relation types.
In contrast, various ontologies (e.g \cite{Hars2003StructureOS,pertsas2017scholarly}) with more than 20 to 35 classes and over 20 to 70 relations and properties are fine-grained and have a relatively high granularity.
\end{itemize}
Although there is usually a correlation between domain specialisation and granularity of the ontology (e.g. an ontology with high domain-specialisation has also a high granularity), there exist also rather domain-independent ontologies with a high granularity, e.g. Scholarly Ontology~\cite{pertsas2017scholarly}), and ontologies with high domain-specialisation and low granularity, e.g. the PICO criterion in Evidence Based Medicine \cite{Kim2011AutomaticCO,Richardson95}) which stands for Population (P), Intervention (I), Comparison (C), and Outcome (O).
Thus, we use both dimensions independently.
Furthermore, a high domain specialisation requirement for a use case implies that each sub-domain requires a separate ontology for the specific use case. These domain-specific ontologies can be organised in a taxonomy.

For the \emph{instance data}, we use the following dimensions: 
\begin{itemize}

\item[C)] \emph{Completeness of the instance data:} 
Given an ontology, to which extent do \emph{all} possible instances (i.e. instances for classes and facts for relation types) in \emph{all} research articles have to be represented in the KG? 
\emph{Low completeness:} it is tolerable for the use case when a considerable amount of instance data is missing for the respective ontology.
\emph{High completeness:} it is mandatory for the use case that for the respective ontology, a considerable amount of instances are present in the instance data.
For instance, given an ontology with a class ``Task'' and a relation type ``subTaskOf'' to describe a taxonomy of tasks, the instance data for that ontology would be complete if all tasks mentioned in all research articles are present (population completeness) and ``subTaskOf'' facts between the tasks are not missing (column completeness).

\item[D)] \emph{Correctness of the instance data:} 
Given an ontology, which correctness is necessary for the corresponding instances? 
\emph{Low correctness:} it is tolerable for the use case, that some instances (e.g. 30\%) are not correct.
\emph{High correctness:} it is mandatory for the use case, that instance data must not be wrong i.e. all present instances in the KG must conform to the ontology and reflect the content of the research articles properly. For instance, an article is correctly assigned to the task addressed in the article, the F1 score in the evaluation results are correctly extracted, etc.
\end{itemize}
It should be noted that completeness and correctness of instance data can be evaluated only for a given ontology. For instance, let A be an ontology having the class ``Deep Learning Model'' without properties, and let B be an ontology that also has a class ``Deep Learning Model'' and additionally further relation types describing the properties of the deep learning model (e.g. drop-out, loss functions, etc.). 
In this example, the instance data of ontology A would be considered to have high completeness, if it covers most of the important deep learning models. 
However, for ontology B, the completeness of the same instance data would be rather low since the properties of the deep learning models are missing.
The same holds for correctness: if ontology B has, for instance, a sub-type ``Convolutional Neural Network'', then the instance data would have a rather low correctness for ontology B if all ``Deep Learning Model'' instances are typed only with the generic class ``Deep Learning Model''.

\subsubsection{Discussion of the KG requirements}

\begin{table*}[tb]
\caption{\textbf{Requirements and approaches for the main use cases.} The upper part describes the minimum requirements for the ontology (domain specialisation and granularity) and the instance data (completeness and correctness). The bottom part 
lists possible approaches for manual, automatic and semi-automatic curation of the KG for the respective use cases. ``X'' indicates that the approach is suitable for the use case while ``(x)'' denotes that the approach is only appropriate with human supervision. The left part (delimited by the vertical triple line) groups use cases suitable for manual, and the right side for automatic approaches. Vertical double lines group use cases with similar requirements.}        
\label{tab:use_cases_approaches}
\begin{tabularx}{\linewidth}{p{1.5cm}|p{3.4cm}|Y|Y||Y|Y|||Y|Y||Y}
                                 &                           & \textit{Extract relevant info} & \textit{Research field overview} & \textit{Deep understanding} & \textit{Repro\-duce results} & \textit{Find related work} & \textit{Recom\-mend articles} & \textit{Assess relevance} \\\hline \hline
\multirow{2}{*}{\textit{Ontology}}        
                                 & Domain specialisation                                   & high                        & high                       & med                   & med            & low               & low              & med  \\\cline{2-9}
                                 & Granularity                                             & high                        & high                       & med                    & med           & low               & low              & low  \\\hline
\multirow{2}{*}{\shortstack[l]{\textit{Instance}\\ \textit{data}}}                                                                                                                         
                                 & Completeness & low                         & med                        & low                       & med        & high              & high             & med     \\\cline{2-9}
                                 & Correctness & med                        & high                       & high                     & high        & low               & low              & med       \\\hline\hline\hline
\multirow{4}{*}{\shortstack[l]{\textit{Manual}\\ \textit{curation}}}                                                                                                                          
                                 & Maintain terminologies                                  & -                            & X                           & -                         & -        & X                 & X                & -             \\\cline{2-9}
                                 & Define templates                                        & X                            & X                           & -                         & -        & -                 & -                & -             \\\cline{2-9}
                                 & Fill in templates                                       & X                            & X                           & X                         & X        & -                 & -                & -             \\\cline{2-9}
                                 & Maintain overviews                                      & X                            & X                           & -                         & -        & -                 & -                & -             \\\hline
\multirow{5}{*}{\shortstack[l]{\textit{Automatic}\\ \textit{curation}}}                                                                                                                        
                                 & Entity/relation extraction                              & (x)                            & (x)                         & (x)                         & (x)  & X                 & X                & X                   \\\cline{2-9}
                                 & Entity linking                                          & (x)                          & (x)                         & (x)                       & (x)      & X                 & X                & X             \\\cline{2-9}
                                 & Sentence classification                                 & (x)                          & -                           & (x)                       & -        & X                 & -                & X             \\\cline{2-9}
                                 & Template-based extraction                               & (x)                          & (x)                         & (x)                         & (x)    & -                 & -                & -                 \\\cline{2-9}
                                 & Cross-modal linking                                     & -                            & -                           & (x)                       & (x)      & -                 & -                & -            
\end{tabularx}

\end{table*}

Next, we discuss the seven main use cases with regard to the required level of ontology domain specialisation and granularity, as well as completeness and correctness of instance data. 
Table~\ref{tab:use_cases_approaches} summarises the requirements for the use cases along the four dimensions at ordinal scale. 
The use cases are grouped together, when they have (1) similar justifications for the requirements, and (2) a high overlap in ontology concepts and instances.

\paragraph{Extract relevant information \& get research field overview:}
The information to be extracted from relevant research articles for a data extraction form within a literature review is very heterogeneous and depends highly on the intent of the researcher and the research questions.
Thus, the ontology has to be domain-specific and fine-grained to offer all possible kinds of desirable information.
However, missing information for certain questions in the KG may be tolerable for a researcher. Furthermore, it is tolerable for a researcher if some of the extracted suggestions are wrong since the researcher can correct them.

Research field overviews are usually the result of a literature review. The data in such an overview has also to be very domain-specific and fine-grained. Also, this information must have high correctness, e.g. an 
F1 score of an evaluation result must not be wrong.
Furthermore, an overview of a particular research field should have appropriate completeness and must not miss any relevant research papers.
However, it is acceptable when 
overviews for some research fields are missing.

\paragraph{Obtain deep understanding \& reproduce results:}
The information required for these use cases has to achieve a high level of correctness (e.g. accurate links to dataset, source code, videos, articles, research infrastructures). 
An ontology for the representation of default artefacts can be rather domain-independent (e.g. Scholix~\cite{Burton_2017}).
However, semantic representation of evaluation protocols require domain-dependent ontologies (e.g. EXPO~\cite{Soldatova2006AnOO}).
Missing information is tolerable for these use cases.

\paragraph{Find related work \& get recommended articles:}
When searching for related work, it is essential not to miss relevant articles. 
Previous studies revealed that more than half of search queries in academic search engines refer to scientific entities~\cite{Xiong2017ExplicitSR}. However, the coverage of scientific entities in general-purpose KGs (e.g. WikiData) is rather low, since the introduction of new concepts in research literature occurs at a faster pace than KG curation~\cite{Ammar2018ConstructionOT}.
Despite the low completeness, Xiong et al. ~\cite{Xiong2017ExplicitSR} could improve the ranking of search results in academic search engines by exploiting general-purpose KGs.
Hence, the instance data for the ``find related work'' use case should have high completeness with fine-grained scientific entities.
However, semantic search engines leverage latent representations of KGs and text (e.g. graph and word embeddings)~\cite{Balog2018EntityOrientedS}.
Since a non-perfect ranking of the search results is tolerable for a researcher, lower correctness of the instance data could be acceptable.
Furthermore, due to latent feature representations, the ontology can be kept rather simple and domain-independent.
For instance, the STM corpus~\cite{Brack2020DomainindependentEO} introduces four domain-independent concepts.

Graph- and content based research paper recommen\-dation systems~\cite{Beel2015ResearchpaperRS} have similar requirements since they also leverage latent feature representations and require fine-grained scientific entities. Also, non-perfect recommendations are tolerable for a researcher.

\paragraph{Assess relevance:}
To help the researcher to assess the relevance of an article according to her needs, the system should highlight the most essential zones in the article to get a quick overview. 
The completeness and correctness of the presented information must not be too low, as otherwise the user acceptance may suffer. 
However, it can be suboptimal, since it is acceptable for a researcher when some of the highlighted information is not essential or when some important information is missing.
The ontology to represent essential information should be rather domain-specific (i.e. using terms that the researchers understands) and quite simple (cf. ontologies for scientific sentence classification in Section~\ref{sec:scientific_information_extraction}).

\section{ORKG construction strategies}

In this section, we discuss the implications for the design and construction of an ORKG and outline possible ap\-proach\-es, which are mapped to the use cases in Table~\ref{tab:use_cases_approaches}.
Based on the discussion in the previous section, we can subdivide the use cases into two groups: 
(1) requiring high correctness and high domain specialisation with rather low requirements on the completeness (left side in Table~\ref{tab:use_cases_approaches}), 
and (2) requiring high completeness with rather low requirements on the correctness and domain specialisation (right side in Table~\ref{tab:use_cases_approaches}).
The first group requires manual approaches while the second group could be accomplished with fully automatic approaches. 
To ensure trustworthiness, data records should contain provenance information, i.e. who or what system curated the data.

\begin{figure}[tb]
    \center{\includegraphics[width=0.9\linewidth]
        {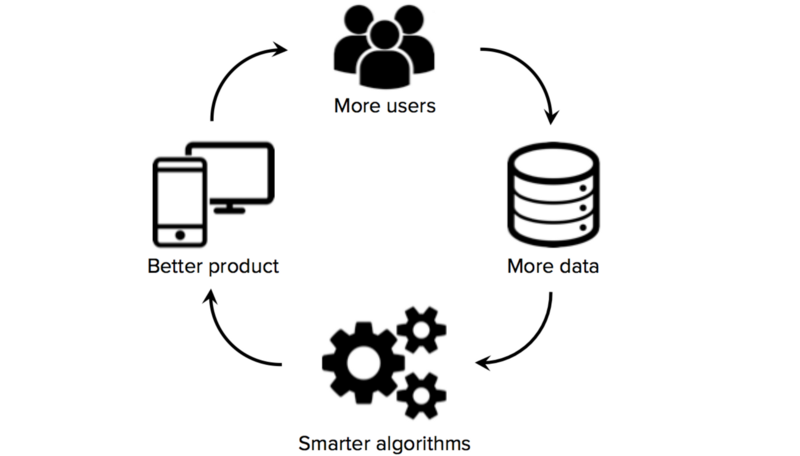}}
    \caption{The virtuous cycle of data network effects by combining manual and automatic data curation approaches \cite{CB_Insights}.}
    \label{fig:data_flywheel}
\end{figure}

Manually curated data can also support use cases with automatic approaches, and vice versa. Furthermore, automatic approaches can complement manual approaches by providing suggestions in user interfaces. Such synergy between humans and algorithms may lead to a ``data flywheel'' (also known as data network effects, see Figure~\ref{fig:data_flywheel}): users produce data which enable to build a smarter product with better algorithms so that more users use the product and thus produce more data, and so on.

\subsection{Manual approaches}

\begin{figure*}[t]
    \center{\includegraphics[width=0.9\linewidth]
        {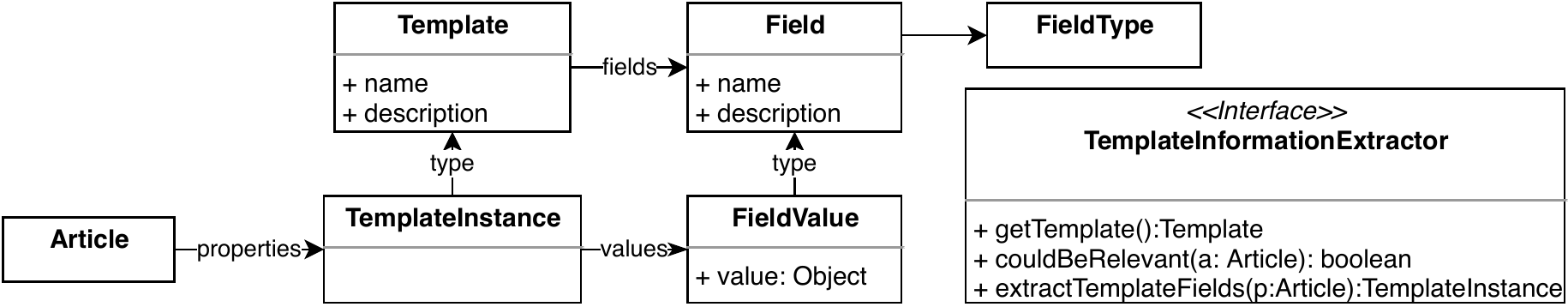}}
    \caption{Conceptual meta-model in UML for templates and interface design for an external template-based information extractor.}
    \label{fig:template_third_party}
\end{figure*}

\paragraph{Ontology design:}
The first group of use cases requires rather domain-specific and fine-grained ontologies. 
We suggest to develop novel or reuse ontologies that fit the respective use case and the specific domain (e.g. EXPO~\cite{Soldatova2006AnOO} for experiments).
Moreover, appropriate and simple user interfaces are necessary for efficient and easy population.

However, such ontologies can evolve with the help of the community, as demonstrated by WikiData and Wikipedia with ``infoboxes'' (see Section~\ref{sec:construction_KG}).
Therefore, the system should enable the maintenance of \emph{templates}, which are pre-defined and very specific forms consisting of fields with certain types (see Figure~\ref{fig:template_third_party}).
For instance, to automatically generate leaderboards for machine learning tasks a template would have the fields Task, Model, Dataset and Score, which can then be filled in by a curator for articles providing such kind of results in a user interface generated from the template. 
Such an approach is based on \emph{meta-modelling}~\cite{Booch2015TheUM}, as the meta-model for templates enables the definition of concrete templates, which are then instantiated for articles.

\paragraph{Knowledge graph population:}
Several user interfaces are required to enable manual population:
(1) populate semantic content for a research article by (1a) choosing relevant templates or ontologies and (1b) fill in the values;
(2) terminology management (e.g. domain-specific research fields);
(3) maintain research field overviews by (3a) assigning relevant research articles to the research field, (3b) define corresponding templates, and (3c) fill in the templates for the relevant research articles.

Furthermore, the system should also offer \emph{Application Programming Interfaces (APIs)} to enable population by third-party applications, e.g.:
\begin{itemize}
 \item Submission portals such as \url{https://www.easychair.org/} during submission of an article.
 \item Authoring tools such as \url{https://www.overleaf.com/} during writing.
 \item Virtual research environments~\cite{Stocker2018TowardsRI} to store evaluation results and links to datasets and source code during experimenting and data analysis.
\end{itemize}
To \emph{encourage crowd-sourced content}, we see the following options:
\begin{itemize}
\item \emph{Top-down enforcement} via submission portals and publishers.
\item \emph{Incentive models}: Researchers want their articles to be cited; semantic content helps other researchers to find, explore and understand an article. This is also related to the  concept of \emph{enlightened self-interest}, i.e. act to further interests of others to serve the own self-interest.
\item Provide \emph{public acknowledgements} for curators.
\item Bring together \emph{experts} (e.g. librarians, researchers from different institutions) who curate and organise content for specific research problems or disciplines.
\end{itemize}

\subsection{(Semi-)automatic approaches}
\paragraph{Ontology design:}
The second group of use cases require a high completeness while a relatively low correctness and domain specialisation are acceptable.
For these use cases, rather simple or domain-independent ontologies should be developed or reused.
Although approaches for automatic ontology learning exist (see Section~\ref{sec:construction_KG}), the quality of their results is not sufficient to generate a meaningful ORKG with complex conceptual models and relations. Therefore, meaningful ontologies should be designed by human experts. 

\paragraph{Knowledge graph population:}
Various approaches can be used to (semi-)automatically populate an ORKG.
Methods for \emph{entity and relation extraction} (see Section~\ref{sec:construction_KG}) can help to populate fine-grained KGs with high completeness and \emph{entity linking} approaches can link mentions in text with entities in KGs.
For cross-modal linking, Singh et al.~\cite{Singh2016OCRAR} suggest an approach to detect URLs to datasets in research articles automatically, while the Scientific Software Explorer \cite{Hoppe2018AnAT} connects text passages in research articles with code fragments.
To extract relevant information at sentence level, approaches for \emph{sentence classification} in scientific text can be applied (see Section~\ref{sec:construction_KG}). 
To support the curator fill in templates semi-automatically, \emph{template-based extraction} can (1) suggest relevant templates for a research article and (2) pre-fill fields of templates with appropriate values. 
For pre-filling, approaches such as n-ary relation extraction~\cite{friedrich-etal-2020-sofc,Hou2019IdentificationOT,Jia2019,Kardas2020-axcell} or end-to-end question answering~\cite{Rajpurkar2016SQuAD10,Devlin2018BERTPO} could be applied.

Furthermore, the system should enable to plugin \emph{external information extractors}, developed for certain scientific domains to extract specific types of information. For instance, as depicted in Figure~\ref{fig:template_third_party}, an external template information extractor has to implement an interface with three methods. This enables the system (1) to filter relevant template extractors for an article and (2) extract field values from an article.

\section{Conclusions}

In this paper, we have presented a requirements analysis for an Open Research Knowledge Graph (ORKG). 
An ORKG should represent the content of research articles in a semantic way to enhance or enable a wide range of use cases.
We identified literature-related core tasks of a researcher that can be supported by an ORKG and formulated them as use cases. 
For each use case, we discussed specificities and requirements for the underlying ontology and the instance data.
In particular, we identified two groups of use cases: 
(1) the first group requires instance data with high correctness and rather fine-grained, domain-specific ontologies, but with moderate completeness;
(2) the second group requires a high completeness, but the ontologies can be kept rather simple and domain-independent, and a moderate correctness of the instance data is sufficient.
Based on the requirements, we have described possible manual and semi-automatic approaches (necessary for the first group), and automatic approaches (appropriate for the second group) for KG construction. 
In particular, we propose a framework with light\-weight ontologies that can evolve by community curation.
Furthermore, we have described the interdependence with external systems, user interfaces, and APIs for third-party applications to populate an ORKG.

The results of our work aim to give a holistic view of the requirements for an ORKG and guide further research. 
The suggested approaches have to be refined, implemented and evaluated in an iterative and incremental process (see \url{www.orkg.org} for the current progress). 
Additionally, our analysis can serve as a foundation for a discussion on ORKG requirements with other researchers and practitioners.


%
\section*{Conflict of interest}
The authors declare that they have no conflict of interest.

\appendix

\section{Comparative Overviews for Information Extraction Datasets from Scientific Text}
\label{sec:comparisons_information_extraction}

Table~\ref{tab:comparison_sentence_classification}, Table~\ref{tab:comparison_relation_extraction}, and Table~\ref{tab:comparison_concept_extraction} show comparative overviews for some datasets from research papers of various disciplines for the tasks sentence classification, relation extraction, and concept extraction, respectively.

\begin{sidewaystable*}%
\footnotesize
\caption{Characteristics of datasets and performance measures for sentence classification in research papers.}
\label{tab:comparison_sentence_classification}
\begin{tabular}{l|l|r|l|l|r|r}
\textbf{Dataset}               & \textbf{Domains}                                                                                  & \textbf{\# Papers} & \textbf{Coverage}  & \textbf{Sentence Classes}                                                                                                                                                                                & \textbf{Inter-coder Agreement}  & \textbf{Performance}               \\ \hline
PubMed-20k     \cite{Dernoncourt2017PubMed2R} & Biomedicine                                                                              & 20,000           & abstracts & \begin{tabular}[c]{@{}l@{}}Background\\ Objective\\ Methods\\ Results, Conclusion\end{tabular}                                                                                                 & n/a                               & 92.9\% F1        \cite{Cohan2019PretrainedLM} \\ \hline
NICTA-PIBOSO   \cite{Kim2011AutomaticCO}      & Biomedicine                                                                              & 1,000            & abstracts & \begin{tabular}[c]{@{}l@{}}Background\\ Intervention\\ Study\\ Population\\ Outcome, Other\end{tabular}                                                                                        & 62.0\% k                          & 84.7\% F1       \cite{Cohan2019PretrainedLM} \\ \hline
CSABSTRUCT     \cite{Cohan2019PretrainedLM}   & Computer Science                                                                         & 2,189            & abstracts & \begin{tabular}[c]{@{}l@{}}Background\\ Objective\\ Method\\ Result, Other\end{tabular}                                                                                                        & 75.0\% k                         & 83.1\% F1       \cite{Cohan2019PretrainedLM}                              \\ \hline
CS-Abstracts   \cite{Goncalves20}             & Computer Science                                                                         & 654              & abstracts & \begin{tabular}[c]{@{}l@{}}Background\\ Objective\\ Methods\\ Results, Conclusions\end{tabular}                                                                                                & n/a                                & 74.6\% F1  \cite{Goncalves20}                              \\ \hline
Emerald 100k   \cite{SteadSBV19}              & \begin{tabular}[c]{@{}l@{}}Management\\ Information Science\\ Engineering\end{tabular}   & 103,457          & abstracts & \begin{tabular}[c]{@{}l@{}}Purpose\\ Design/methodology/approach\\ Findings\\ Originality/value\\ Social implications\\ Practical implications\\ Research limitations/implications\end{tabular} & n/a                           & n/a                                         \\ \hline
MAZEA          \cite{Dayrell2012RhetoricalMD} & \begin{tabular}[c]{@{}l@{}}Physics\\ Engineering\\ Life and Health Sciences\end{tabular} & 1,335            & abstracts & \begin{tabular}[c]{@{}l@{}}Background\\ Gap, Purpose\\ Method\\ Result, Conclusion\end{tabular}                                                                                               & 59,4\% k            &    66.0\% accuracy \cite{Dayrell2012RhetoricalMD} \\ \hline                              
Safder et al.  \cite{Safder2020}              & Computer Science                                                                         & 92               & full text & \begin{tabular}[c]{@{}l@{}}Algorithmic Efficiency\\ Dataset Description\\ Algorithmic Time Complexity\\ Other\end{tabular}                                                                      & n/a                              & 78.5\% accuracy \cite{Safder2020}                               \\ \hline
Dr. Inventor   \cite{Fisas2015OnTD}           & Computer Graphics                                                                        & 40               & full text & \begin{tabular}[c]{@{}l@{}}Background\\ Challenge\\ Approach\\ Outcome, Future Work\end{tabular}                                                                                               & 66.7\% k                          & 72.5\% accuracy \cite{Badie2018ZoneIB}       \\ \hline
ART/CoreSC     \cite{Liakata2010CorporaFT}    & \begin{tabular}[c]{@{}l@{}}Chemistry\\ Computational Linguistic\end{tabular}             & 225              & full text & \begin{tabular}[c]{@{}l@{}}Background\\ Motivation, Goal\\ Hypothesis\\ Object\\ Model, Method\\ Experiment, Result\\ Observation, Conclusion\end{tabular}                                  & 57.0\% k                         & 51.6\% F1       \cite{liakata2012automatic}  
\end{tabular}
\end{sidewaystable*}

\begin{sidewaystable*}
\footnotesize
\caption{Characteristics of datasets and performance measures for binary and n-ary relation extraction in research papers. \textsuperscript{*} For SOFC-Exp corpus, performance values were obtained with ground truth concept mentions.}
\label{tab:comparison_relation_extraction}
\begin{tabular}{l|l|r|l|l|l|l|l|r|r}
\textbf{Dataset}   & \textbf{Domains}                                                                                & \textbf{\# Papers} & \textbf{Coverage}  & \textbf{Cardinality} & \textbf{Relation Types} & \textbf{Scope} & \textbf{Inter-coder Agreement} & \textbf{\# Relations} & \textbf{Performance}                \\ \hline
SemEval17         \cite{augenstein2017semeval}    & \begin{tabular}[c]{@{}l@{}}Computer Science\\ Material Sciences\\ Physics\end{tabular} & 500              & abstract  & binary               & \begin{tabular}[c]{@{}l@{}}synonym-of\\ hyponym-of\end{tabular}                                                                                                                                                                                                                & intra-sentence  & 60.0\% k               & 672              & 28.0\% F1      \cite{augenstein2017semeval}    \\ \hline
SemEval18        \cite{Zadeh2016TheAR}           & Comp. Linguistics                                                              & 500              & abstract  & binary               & \begin{tabular}[c]{@{}l@{}}usage\\ result\\ model\\ part-whole\\ topic\\ comparison\end{tabular}                                                                                                                                                                               & intra-sentence  & 90.8\% F1              & 1595             & 49.3\% F1      \cite{Zadeh2016TheAR}           \\ \hline
ChemProt          \cite{Kringelum2016}            & Biomedicine                                                                            & 2482             & abstract  & binary               & \begin{tabular}[c]{@{}l@{}} UPREGULATOR \\ACTIVATOR \\ DOWNREGULATOR \\ INHIBITOR \\ AGONIST\\ANTAGONIST\\ SUBSTRATE\\ \end{tabular} & intra-sentence  & n/a                   & 10,031           & 83.64\% F1    \cite{Beltagy2019SciBERTPC}     \\ \hline
SciERC            \cite{Luan2018MultiTaskIO}      & Art. Intelligence                                                                 & 500              & abstract  & binary               & \begin{tabular}[c]{@{}l@{}}hyponym-of\\ compare\\ part-of\\ conjunction\\ evaluate-for\\ feature-of\\ used-for\end{tabular}                                                                                                                                                    & cross-sentence  & 67.8\% F1              & 4,716             & 39.3\% F1      \cite{Luan2018MultiTaskIO}      \\ \hline
PWC  \cite{Kardas2020-axcell}        & Art. Intelligence                                                                 & 731              & full text & n-ary                & \begin{tabular}[c]{@{}l@{}}(Task,\\Dataset,\\Metric,\\Score)\end{tabular}                                                                                                                                                                                                                                                 & document-level  & n/a                   & 2,295            & 28.7\% F1      \cite{Kardas2020-axcell}        \\ \hline
CKB               \cite{Jia2019}                  & Biomedicine                                                                            & 343              & full text & n-ary                & \begin{tabular}[c]{@{}l@{}}(Drug, Gene, \\Mutation)\end{tabular}                                                                                                                                                                                                                                                            & document-level  & n/a                   & 2,025            & 52.8\% F1      \cite{Jia2019}                  \\ \hline
SOFC-Exp          \cite{friedrich-etal-2020-sofc} & Material Sciences                                                                      & 45               & full text & n-ary                & \begin{tabular}[c]{@{}l@{}}(AnodeMaterial, \\ CathodeMaterial, \\ Device, \\ ElectrolyteMaterial, \\ FuelUsed, \\ InterlayerMaterial, \\ OpenCircuitVoltage, \\ PowerDensity, \\Resistance, \\ WorkingTemperature)\end{tabular}                                                        & document-level  & n/a                   & n/a              & 56.4\% F1\textsuperscript{*}      \cite{friedrich-etal-2020-sofc}
\end{tabular}
\end{sidewaystable*}

\begin{sidewaystable*}
\footnotesize
\caption{Characteristics of datasets and performance measures for scientific concept extraction in research papers. \textsuperscript{*} For SOFC-Exp corpus, performance values were obtained with ground truth sentences describing experiments.}
\label{tab:comparison_concept_extraction}
\begin{tabular}{l|l|r|r|l|l|r|r}
\textbf{Dataset}                        & \textbf{Domains}                                                                                                                                                                                                                 & \textbf{\# Papers} & \textbf{\# Concepts} & \textbf{Coverage}  & \textbf{Concept Types}                                                                                                                         & \textbf{Inter-coder Agreement} & \textbf{Performance}                    \\ \hline
SemEval17     \cite{augenstein2017semeval}        & \begin{tabular}[c]{@{}l@{}}Computer Science\\ Material Sciences\\ Physics\end{tabular}                                                                                                                                  & 500              & 9,946           & abstract  & \begin{tabular}[c]{@{}l@{}}Process\\ Task\\ Material\end{tabular}                                                                     & 60.0\% $\kappa$               & 56.9\% F1	\cite{park-caragea-2020-scientific}          \\ \hline
STM           \cite{Brack2020DomainindependentEO} & \begin{tabular}[c]{@{}l@{}}Agriculture \\ Astronomy  \\ Biology  \\ Chemistry \\ Computer Science  \\ Earth Science  \\ Engineering \\ Materials Science \\ Mathematics \\ Medicine\end{tabular} & 110              & 6,127           & abstract  & \begin{tabular}[c]{@{}l@{}}Process\\ Method\\ Material\\ Data\end{tabular}                                                            & 76.0\% $\kappa$               & 65.5\% F1      \cite{Brack2020DomainindependentEO} \\ \hline
SciERC        \cite{Luan2018MultiTaskIO}          & Art. Intelligence                                                                                                                                                                                                  & 500              & 8,089           & abstract  & \begin{tabular}[c]{@{}l@{}}Task\\ Method\\ Metric\\ Material\\ Other\\ Generic\end{tabular}                                           & 76.9\% $\kappa$               & 75.2\% F1	\cite{park-caragea-2020-scientific}        \\\hline
ACL2          \cite{Zadeh2016TheAR}               & Comp. Linguistics                                                                                                                                                                                               & 300              & 6,818           & abstract  & \begin{tabular}[c]{@{}l@{}}Method\\ Tool\\ Language Resource (LR)\\ LR product\\ Model\\ Measures/Measurements\\ Other\end{tabular} & 63.0\% F1              & 69.9\% F1	\cite{park-caragea-2020-scientific}    \\ \hline
B5CDR         \cite{Li2016}                       & Biomedicine                                                                                                                                                                                                             & 1500             & 28,785          & abstract  & \begin{tabular}[c]{@{}l@{}}Chemical\\ Disease\end{tabular}                                                                            & 91.8\% F1               & 88.9\% F1    \cite{Beltagy2019SciBERTPC}         \\ \hline
NCBI-disease  \cite{Dogan2014}                    & Biomedicine                                                                                                                                                                                                             & 793              & 6,892           & abstract  & Disease                                                                                                                               & 88.0\% F1               & 96.9\% F1     \cite{Beltagy2019SciBERTPC}  \\ \hline
SOFC-Exp      \cite{friedrich-etal-2020-sofc}     & Material Sciences                                                                                                                                                                                                       & 45               & 4,004           & full text & \begin{tabular}[c]{@{}l@{}}Material\\ Device\\ Value\end{tabular}                                                                     & 95.8\% F1             & 81.5\% F1\textsuperscript{*}      \cite{friedrich-etal-2020-sofc}     \\ \hline
\end{tabular}
\end{sidewaystable*}

\bibliographystyle{spmpsci}      
\bibliography{references}   

%
%

\end{document}